\documentclass[10pt,aps,prl,twocolumn,english]{revtex4-1}

\usepackage{graphicx}
\usepackage{amsmath}
\usepackage{amssymb}
\usepackage{dsfont}
\usepackage{fancyhdr}
\usepackage{yfonts}
\usepackage{braket}
\usepackage{siunitx}
\usepackage{booktabs}

\usepackage[utf8]{inputenc}
\DeclareFontFamily{U}{BOONDOX-calo}{\skewchar\font=45 }
\DeclareFontShape{U}{BOONDOX-calo}{m}{n}{
	<-> s*[1.05] BOONDOX-r-calo}{}
\DeclareFontShape{U}{BOONDOX-calo}{b}{n}{
	<-> s*[1.05] BOONDOX-b-calo}{}
\DeclareMathAlphabet{\mathcalboondox}{U}{BOONDOX-calo}{m}{n}
\SetMathAlphabet{\mathcalboondox}{bold}{U}{BOONDOX-calo}{b}{n}
\DeclareMathAlphabet{\mathbcalboondox}{U}{BOONDOX-calo}{b}{n}

\usepackage{hyperref}
\hypersetup{
	colorlinks   = true, 
	urlcolor     = blue, 
	linkcolor    = blue,
	citecolor   = blue,
}

\begin{document}

\title[]{Non-adiabatic Storage of Short Light Pulses in an Atom--Cavity System}

\author{Tobias~\surname{Macha}}\email{macha@iap.uni-bonn.de}
\author{Eduardo~\surname{Uru{\~n}uela}}
\author{Wolfgang~\surname{Alt}}
\author{Maximilian~\surname{Ammenwerth}}
\author{Deepak~\surname{Pandey}}
\author{Hannes~\surname{Pfeifer}}
\author{Dieter~\surname{Meschede}}
\affiliation{Institut f{\"u}r Angewandte Physik, Universit{\"a}t Bonn, Wegelerstra{\ss}e 8, 53115~Bonn, Germany}

\begin{abstract}
We demonstrate the storage of $5$~ns light pulses in a single rubidium atom coupled to a fiber-based optical resonator. Our storage protocol addresses a regime beyond the conventional adiabatic limit and approaches the theoretical bandwidth limit. We extract the optimal control laser pulse properties from a numerical simulation of our system and measure storage efficiencies of $(8.2\pm 0.6)~\%$, in close agreement with the maximum expected efficiency. Such well-controlled and high-bandwidth atom-photon interfaces are key components for future hybrid quantum networks.
\end{abstract}

\maketitle

Quantum networks are the basis for distributed quantum information processing~\cite{Kok2007,Cirac1999,Sangouard2011} and long-distance quantum communication~\cite{Briegel1998,Duan2001}. In quantum networks~\cite{Cirac1997}, distant nodes are connected via quantum channels, e.g. optical fibers guiding single photons as flying qubits~\cite{Northup2014}.
The nodes for processing and storage of quantum information require long coherence times and the ability to efficiently convert flying to stationary qubits and vice versa. Single atoms in optical cavities have shown to fulfill these criteria~\cite{Koerber2018,Boozer2007}, but so far only in the adiabatic regime of atom-cavity dynamics when interacting with photon pulses of length $T\gg\kappa/g^2$, where $\kappa$ is the cavity bandwidth and $g$ is the atom-cavity coupling strength.
Both working in this regime and the choice of cavity parameters have limited previous experiments to pulses much longer than the atomic excited state lifetime $\tau_e$ and the cavity field decay time~\cite{Boozer2007,Specht2011}. However, high-bandwidth quantum communication will use short pulses, such as the polarization-entangled photons emitted by quantum dots~\cite{Keil2017} or spontaneous parametric down-conversion sources~\cite{Burnham1970}.

In our approach, we use a high-bandwidth, microscopic fiber Fabry-Pérot cavity (FFPC)~\cite{Hunger2010}, strongly coupled to a single atom, 
to store a weak coherent pulse in the non-adiabatic regime near $T\sim\kappa/g^2$~\cite{Gorshkov2007}. This way, pulses with $T\ll\tau_e$ are stored, which is not possible in free space~\cite{Gorshkov2007b,Steiner2017}. The cavity is thus used as a bandwidth converter, matching the narrow atomic transition to a spectrally broad pulse near the cavity-bandwidth limit $T^{-1}\sim\kappa$. The pulse is mapped into the atomic ground states with the help of a control laser in Raman configuration. In order to execute the storage process efficiently, the exact control pulse properties, such as the temporal profile, have to be found and matched to the input pulse.
Prominent theoretical work~\cite{Fleischhauer2000,Gorshkov2007,Dilley2012} has been mostly concerned with realizing an adiabatic state transfer during the storage process and hence is not applicable in our case. Instead, we determine an optimum pulse sequence in the non-adiabatic storage regime by numerical simulations based on the full quantum-mechanical Lindblad master equation describing our system. As a result, we reach an excellent agreement between expected and measured storage efficiencies.

\begin{figure*}[ht]
	\includegraphics[width=1\textwidth]{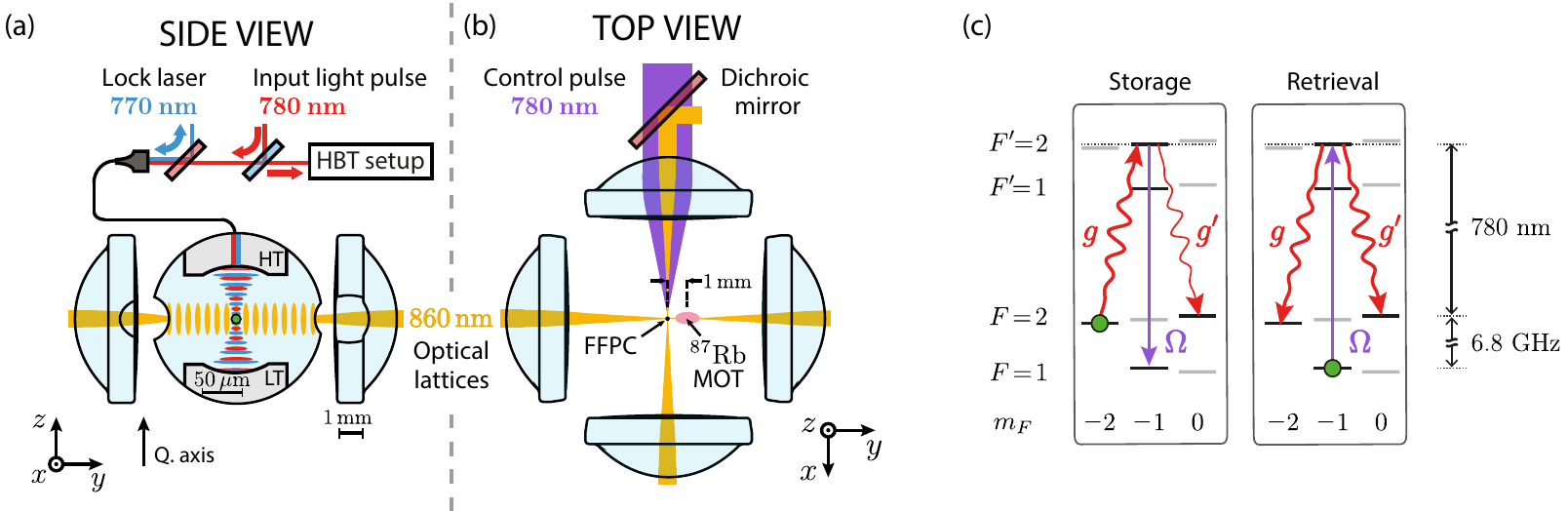}
	\caption[Setup]{Schematic side (a) and top (b) views of the experimental setup illustrate the optical lattices trapping a single rubidium ($^{87}$Rb) atom at the center of a microscopic fiber Fabry-Pérot cavity (FFPC). The high transmission (HT) mirror of the cavity is the access port for a coherent input light pulse, which is stored in the atomic memory via a control pulse entering from the side. Retrieved single photons are guided to a Hanbury Brown-Twiss (HBT) setup for detection. (c)~Photon storage for a cavity with two degenerate polarization modes. The first mode couples the initial state $\ket{F,m_F}=\ket{2,-2}$ to the excited state $\ket{2',-1}$ with a rate $g$ and either results in coherent transfer to $\ket{1,-1}$ by the interaction with the control laser (Rabi frequency $\Omega$) or in coherent leakage to $\ket{2,0}$ via a second cavity mode with rate $g'$. The efficiency of an adiabatic retrieval process is not affected as the photon detection is polarization-insensitive.
	}
	\label{fig:Setup}
\end{figure*}

Our photon memory consists of a single $^{87}$Rb atom trapped at the center of a single-sided, high-bandwidth FFPC~\cite{Gallego2018}. One of the fiber mirrors presents a higher transmission (HT), ensuring a highly directional input-output channel~\cite{Gallego2016}. As depicted in Fig.~\ref{fig:Setup}a, the cavity is placed at the focus of four in-vacuum, aspheric lenses (NA= 0.5), which lead to a high beam pointing stability~\cite{Dorantes2017}. The lenses strongly focus two pairs of counter-propagating, red-detuned dipole trap beams at 860~nm which create a 2D optical lattice in the $xy$-plane, see Fig.~\ref{fig:Setup}b. One of the lattices acts as a conveyor belt~\cite{Kuhr2001} to transport atoms from a magneto-optical trap (MOT) into the cavity. Confinement in the $z$-direction is provided by the intra-cavity, blue-detuned lock laser field at 770 nm, which is additionally used for stabilizing the resonator length and for carrier-free Raman cooling in three dimensions~\cite{Reimann2014a,Neuzner2018}. As a result, the atom is located with sub-wavelength precision at an antinode of the cavity mode driven by the input pulse. In particular, the mode is resonant to the Stark-shifted $\ket{F=2,m_F=-2}\rightarrow\ket{F'=2,m_F=-1}$ hyperfine transition of rubidium at $780$~nm. The quantization axis is aligned with the cavity axis by applying a magnetic guiding field of $\sim1.8$~G.

In each experimental cycle, the memory is initialized by cooling and preparing the atom in the state $\ket{F=2,m_F=-2}$ by optical pumping with an efficiency exceeding $95~\%$. As a first step of the storage protocol, we send a triggered, coherent input pulse with a mean photon number $n$ and a duration of $5$~ns (FWHM). It has a time-symmetric, sine-squared shaped probability amplitude $|\phi_{\text{in}}|(t)$ of the electric field~\cite{Cirac1997}. When it enters the FFPC through the HT mirror, a classical control laser pulse in two-photon resonance is simultaneously applied from the side along the $x$-axis (Fig.~\ref{fig:Setup}b). This results in transferring the atom dominantly to the state $\ket{F=1, m_F=-1}$, see Fig.~\ref{fig:Setup}c.
After a storage time of $1~\mu$s, the photon is read out with an adiabatic control pulse to ensure maximum population transfer~\cite{Mucke2013,Nisbet-Jones2011,Keller2004}. The cycle of state initialization, photon storage and retrieval is repeated with the same atom up to 1500 times for $\sim2$ seconds, limited by the efficiency of the currently employed cooling mechanism.

To find the optimum storage-assisting control laser pulse with time-dependent Rabi frequency $\Omega(t)$, we simulate the system based on a Lindblad master equation. The underlying Hamiltonian consists of the Jaynes-Cummings Hamiltonian~\cite{Shore1993} and an additional driving term
\begin{equation*}
\begin{aligned}
\hat{H}_{\text{d}}(t)=\,&i\,\hbar\,\frac{\Omega(t)}{2}\left(\hat{\sigma}^\dagger-\hat{\sigma}\right)\\
&+\hbar\,\sqrt{2\kappa_{\text{HT}}}\cdot\sqrt{n}\cdot\phi_{\text{in}}(t)\left(\hat{a}^\dagger+\hat{a}\right)\,
\end{aligned}
\end{equation*}
where $\hat{\sigma}^\dagger,\hat{\sigma}$ are the flip operators of the atomic states that are coupled via $\Omega(t)$, while $\hat{a}^\dagger(\hat{a})$ is the creation (annihilation) operator of the driven cavity mode. The total cavity damping $\kappa=\kappa_{\text{HT}}+\kappa_{\text{loss}}$ is the sum of the pure transmission rate $\kappa_{\text{HT}}$, at which a coherent field ($\phi_{\text{in}}(t)$) impinging on the HT mirror interacts with the open system, and the undesired losses $\kappa_{\text{loss}}$, e.g. due to absorption and scattering on the mirrors (for more details see Supplemental Material).

\begin{figure*}[ht]
	\includegraphics[width=1\textwidth]{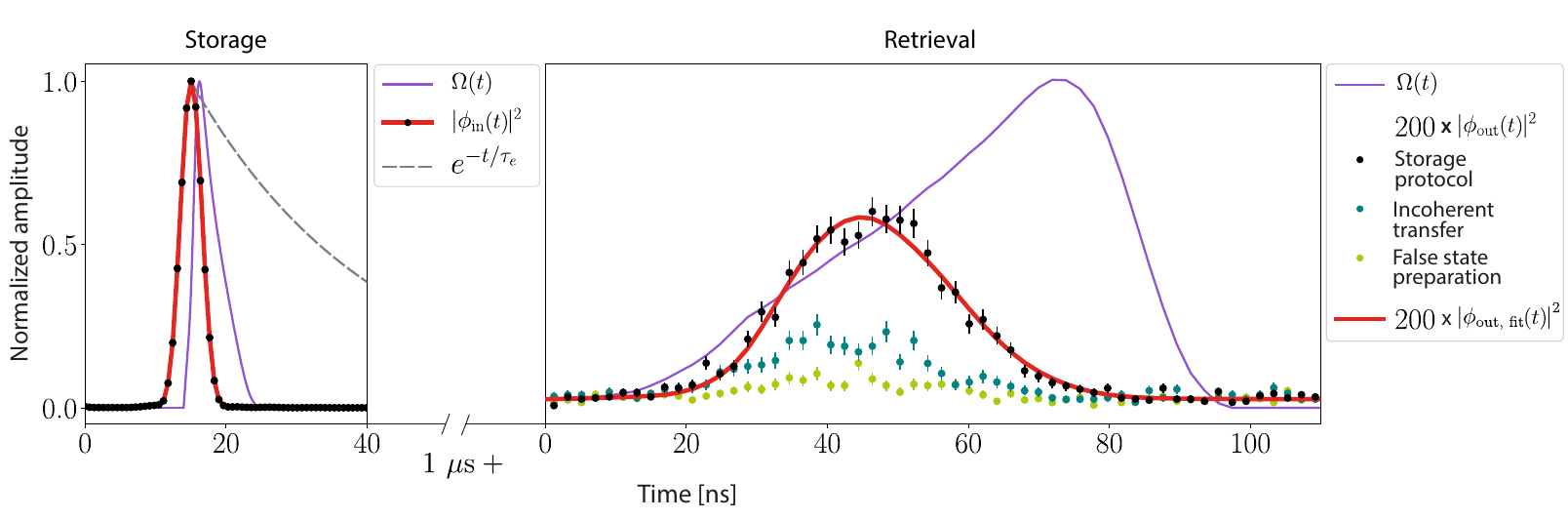}%
	\caption{On the left: The time-symmetric, sine-squared shaped input pulse with intensity probability amplitude $\left|\phi_{\text{in}}(t)\right|^2$ sent to the HT mirror has a FWHM duration of $5$~ns (black points connected by red, solid line). For comparison, the atomic excited state decay with a time constant of $\tau_e=26$~ns is shown (gray, dashed line). The Raman control pulse with Rabi frequency $\Omega(t)$ (purple, solid line) is applied with a delay of $4$~ns with respect to the input pulse, in contrast to adiabatic storage protocols (see main text). All pulse amplitudes are normalized to one. On the right: After a storage time of $1~\mu$s, a control pulse (purple, solid line) adiabatically generates a photon $\left|\phi_{\text{out}}(t)\right|^2$ after the full storage protocol (black dots). By taking into account the incoherently transferred population in the absence of a control pulse (blue dots) and the counts due to false initial state preparation (green dots), we infer a coherent storage component of $(79\pm 3)~\%$. The data point values have been scaled by a factor of 200, while the Raman pulse is still normalized to one. From a simulation-based fit $\left|\phi_{\text{out, fit}}(t)\right|^2$ (red, solid line) we extract the atom-cavity coupling strengths $(g,g')=2\pi\cdot (29,35)$~MHz.
	}
	\label{fig:storage}
\end{figure*}
\begin{figure*}[ht]
	\includegraphics[width=1\textwidth]{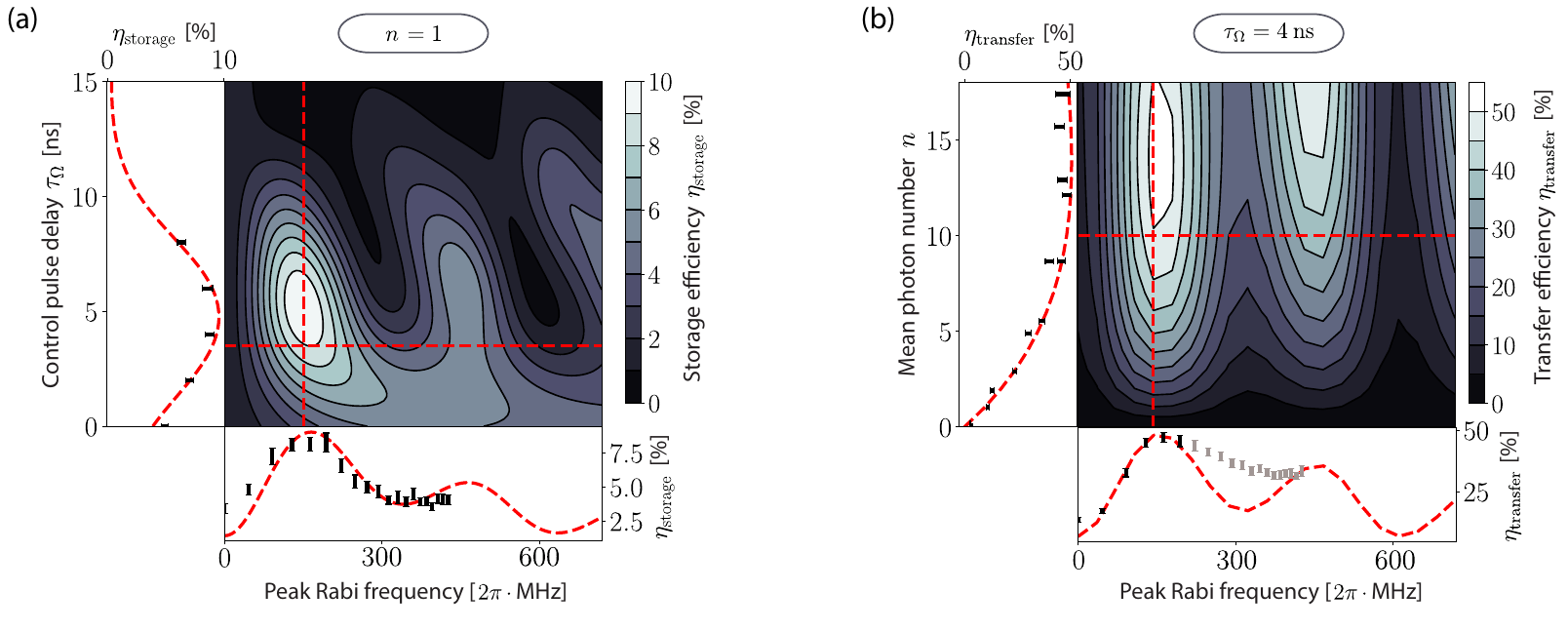}
	\caption{Simulated efficiency maps and experimental pulse parameter scans.~(a)~For an input pulse containing a mean photon number of $n=1$, the storage efficiency $\eta_{\text{storage}}$ as a function of both control pulse peak Rabi frequency and control pulse delay $\tau_{\Omega}$ is simulated. The non-adiabatic storage process reveals a significant atomic excited state population, which is then mapped by the control laser to the target ground state in a coherent $\pi$-pulse process. Thus, for higher peak Rabi frequencies efficiency revivals are observed. (b)~For a fixed control pulse delay of $4$~ns, the transfer efficiency $\eta_{\text{transfer}}$ is simulated as a function of both control pulse peak Rabi frequency and mean photon number per input pulse $n$.
	Four independent, experimental parameter scans (black points) are fitted to the simulation of our system (red, dashed lines).
	This allows to extract the storage efficiency $\eta_{\text{storage}}=(8.2\pm0.6)~\%$ for a coherent input pulse with a mean photon number of $n=1$. Data points in gray are not considered by the fit (see main text).
}
	\label{fig:efficiency}
\end{figure*}

Our cavity supports two degenerate polarization modes ($\sigma^\pm$), which couple two Zeeman states in the same hyperfine manifold $F=2$ via the excited state $\ket{2',-1}$ (Fig.~\ref{fig:Setup}c). With the $\pi$-polarized control laser coupling the excited state to the $F=1$ manifold, our choice of the initial Zeeman state leads to coherent dynamics in a tripod configuration~\cite{Vitanov2017}. Additionally, the probability to off-resonantly excite the state $\ket{1',-1}$ has to be considered. We take all of these effects into account in our model by including two polarization cavity modes with effective atom-cavity coupling strengths $g,g'$ and a total of five atomic states (for a detailed discussion see Supplemental Material). The main effect of the ideally absent atom-cavity coupling $g'$ is a coherent population leakage during a storage attempt. The photon storage efficiency $\eta_{\text{storage}}$, which is the transfer efficiency $\eta_{\text{transfer}}$ from initial to target state normalized by the mean input photon number ($\eta_{\text{storage}}=\eta_{\text{transfer}}/n$), is thus decreased compared to a standard Lambda configuration~\cite{Gorshkov2007}.

In general, the storage efficiency depends crucially on the properties of the control pulse, namely the temporal shape, the pulse amplitude, the detuning from the atomic transition and the delay with respect to the input pulse. However, in a non-adiabatic regime we find that its exact temporal shape plays a minor role. A simple compression of the temporal length of the pulse shape for the adiabatic protocol~\cite{Dilley2012} is equally effective as a numerically optimized pulse shape for our short input pulse~\cite{Giannelli2018}. In case of zero single-photon detuning of the input pulse with respect to the atomic excited state, our simulation predicts the highest storage efficiency for a vanishing two-photon detuning of the Raman transition, as also predicted by~\cite{Gorshkov2007}. The remaining pulse parameters for optimal storage are the peak Rabi frequency of the control laser and its delay $\tau_{\Omega}$ with respect to the input light pulse.

Besides the pulse parameters, knowledge about the system parameters $g,g',\kappa_{\text{HT}},\kappa_{\text{loss}},\gamma$ is important for the storage process. $\kappa_{\text{HT}}$ is known from the mirror characterization in~\cite{Gallego2016} and $\kappa_{\text{loss}}=\kappa-\kappa_{\text{HT}}$ is obtained after measuring $\kappa$ by probing the frequency-dependent cavity reflection. For determining $g,g'$, we take a measurement, during which we store an input pulse with on average $n=2.1$ photons and reconstruct the retrieved pulse after the memory read-out, as shown in Fig.~\ref{fig:storage}. A simulation-based fit of the resulting shape with $g,g'$ as free parameters completes our set of system parameters: $(g,g',\kappa_{\text{HT}},\kappa_{\text{loss}},\gamma)=2\pi\cdot(29,35,16,25,3)$~MHz.

Two additional measurements determine the coherent storage fraction. The first one omits the control pulse during the storage process and thus indicates the incoherent state transfer due to optical pumping by the input pulse. The second measurement uses neither control nor input pulse, which indicates false state preparation. From the ratio of the integrated detection counts in Fig.~\ref{fig:storage}, we obtain a coherent storage component of $(79 \pm3)~\%$ (for the special case of $n=2.1$). In a Hanbury Brown-Twiss experiment we verify the single-photon character of the retrieved pulses by calculating the correlation function $g^{(2)}(0)=(12\pm6)~\%$ from the time trace of detected photons. The value is consistent with the amount of background light and detector dark counts (see Supplemental Material).

In a next step, the previously obtained system parameters are used to simulate the storage process in order to map out the full parameter space for the optimization of the storage efficiency $\eta_{\text{storage}}$. In Fig.~\ref{fig:efficiency}a, $\eta_{\text{storage}}$ is displayed as a function of the peak Rabi frequency of the control laser and its delay $\tau_{\Omega}$ and in Fig.~\ref{fig:efficiency}b, the transfer efficiency $\eta_{\text{transfer}}$ as a function the peak Rabi frequency and mean photon number per input pulse is shown. The latter is of interest for cross-checking the photon number calibration, which is required to determine the storage efficiency rather than the transfer efficiency.

The simulation results show efficiency revivals towards higher Rabi frequencies, which give insight into the underlying storage process. The revivals are a consequence of the excited state being significantly populated before it is mapped by the control laser to the target ground state in a coherent $\pi$-pulse interaction~\cite{Gorshkov2007}. In contrast, a classic STIRAP protocol~\cite{Shore2017} does not show revivals. It relies on the adiabatic transfer between the ground states, which is no longer the most efficient storage method in the presented experiment.

We confirm the simulated behavior by measuring four independent parameter scans, which are fitted to the experimentally accessible regions of the two simulated maps. To obtain the storage and transfer efficiencies from the measurement, the photon detection probabilities per storage attempt are corrected for the imperfect state preparation, the read-out efficiency of $(80\pm5)~\%$, the transmission in the optical path and the detection efficiencies of $(19\pm 7)~\%$ and the spatial mode matching between fiber-guided and cavity mode of $(60\pm 2)~\%$~\cite{Gallego2016}.
For an input pulse with $n = 1$ (see Fig.~\ref{fig:efficiency}a) we observe a maximum while scanning the control laser peak Rabi frequency, from which we deduce the storage efficiency of $\eta_{\text{storage}}=(8.2\pm 0.6)~\%$, which is close to the highest expected value of $9.0~\%$ for our tripod system with cavity losses. Taking the aforementioned efficiencies into account, the end-to-end efficiency of creating an outgoing single-photon Fock state per impinging coherent state is $(0.9\pm0.1)~\%$. For a larger mean photon numbers per pulse (see Fig.~\ref{fig:efficiency}b), we observe the expected saturation of the transfer efficiency, which is limited by the undesired transfer to $\ket{2,0}$ (see also Supplemental Material). 
However, for higher peak Rabi frequencies the measured data deviates from the simulation (Fig.~\ref{fig:efficiency}b).
We attribute this behavior to variations in both the atom-cavity coupling strength and the ac Stark shift, which originate from different atom positions within the cavity mode and dipole traps. As a result, the optimum (two-photon) Rabi frequency is met at higher peak Rabi frequencies than expected, leading to the observed higher efficiencies.

With technical improvements such as the realization of a three-level (Lambda) configuration, the efficiency can already be improved by a factor of two. Assuming negligible undesired cavity losses, storage efficiencies exceeding $40~\%$ should be feasible with a single atom. The overall memory efficiency can be increased by fiber cavities equipped with GRIN lenses~\cite{Gulati2017}, which reduce the losses due to a cavity-fiber mode-mismatch.

In conclusion, we have demonstrated the non-adiabatic storage of light pulses, which are, with $5$~ns, much shorter than the atomic excited state lifetime of $\tau_e=26$~ns. By simulating the storage process in dependence of the control pulse parameters, we find the optimum control pulse for the highest possible photon storage efficiency and -- for the first time -- observe a remarkable agreement with experimentally obtained values.

Our system is capable of interacting with very short light pulses in a highly directional manner, thereby demonstrating functionality for a high-bandwidth quantum network. Additionally, FFPCs offer an intrinsic fiber coupling that facilitates the implementation in cavity-based networks~\cite{Kimble2008,Reiserer2015}. In the future, we will employ ensembles of atoms, which will enhance the light-matter interaction by collective effects~\cite{Thompson1992}, allowing for storage of even shorter pulses with even higher bandwidths. In this way, true single-photon Fock states as provided by the emission of a quantum dot~\cite{Keil2017} can efficiently interact with our atom-based system. Envisioning such a hybrid experiment~\cite{Meyer2015}, we have recently demonstrated that the emission frequency of quantum dots can be stabilized to atomic transitions~\cite{ZopfMacha2018}.

\bigskip
This work has been funded by the Deutsche Forschungsgemeinschaft (DFG, German Research Foundation) - Project number 277625399- TRR 185 and the Bundesministerium für Bildung und Forschung (BMBF), project Q.Link.X. We thank J. Gallego and E. Keiler for discussions and technical support in the early stage of the presented experiment.


\bibliography{Biblio}
\bibliographystyle{apsrev4-1}
\end{document}